\documentclass[traditabstract]{aa}
\usepackage{graphicx}
\usepackage{natbib}
\usepackage{txfonts}
\bibpunct{(}{)}{;}{a}{}{,}
\begin{document}

  \title{On the missing second generation AGB stars in NGC~6752}
  
  \author{Santi Cassisi\inst{1},  Maurizio Salaris\inst{2}, Adriano Pietrinferni\inst{1}, Jorick S. Vink\inst{3}, \and Matteo Monelli\inst{4}}

\institute{INAF~$-$~Osservatorio Astronomico di Collurania, Via M. Maggini, I$-$64100 , Teramo, Italy, 
            \email{cassisi,pietrinferni@oa-teramo.inaf.it} 
           \and Astrophysics Research Institute, 
           Liverpool John Moores University, 
           IC2, Liverpool Science Park, 
           146 Brownlow Hill, 
           Liverpool L3 5RF, UK, \email{M.Salaris@ljmu.ac.uk}           
           \and
           Armagh Observatory, College Hill, Armagh BT61 9DG, Northern Ireland, \email{jsv@arm.ac.uk}
           \and
           Instituto de Astrofisica de Canarias, Calle Võa L{\'a}ctea s/n, E-38205 La Laguna, Tenerife, Spain,
           \email{monelli@iac.es}
           }

 \abstract{In recent years the view of Galactic 
globular clusters as simple stellar populations has changed dramatically, as it  
is now thought that basically all globular clusters host multiple stellar 
populations, each with its own chemical abundance pattern and colour-magnitude diagram sequence.
Recent spectroscopic observations of asymptotic giant branch stars in  
the globular cluster NGC\,6752 have disclosed a low [Na/Fe] abundance for the whole 
sample, suggesting that they are all 
first-generation stars, and that all second-generation stars fail to reach the AGB in this cluster.
A scenario proposed to explain these observations invokes strong mass loss in second-generation 
horizontal branch stars --all located at the hot side of the blue and extended 
horizontal branch of this cluster-- possibly induced by the metal enhancement associated to 
radiative levitation. This enhanced mass loss would prevent second generation stars from  
reaching the asymptotic giant branch phase, thus explaining at the same time the low value 
of the ratio between horizontal branch and asymptotic giant branch stars (the ${\rm R_2}$ parameter) 
observed in NGC6752. We have critically discussed this mass-loss scenario, finding 
that the required mass-loss rates are of the order of
$10^{-9} {\rm M_{\odot} yr^{-1}}$, significantly higher than
current theoretical and empirical constraints. 
By making use of synthetic horizontal branch simulations, we demonstrate that our modelling  
predicts correctly the ${\rm R_2}$ parameter for NGC6752,  
without the need to invoke very efficient mass loss during the core He-burning stage. 
As a test of our stellar models we show that we can  
reproduce the observed value of ${\rm R_2}$ for both M3, a cluster of approximately the same metallicity and with 
a redder horizontal branch morphology, and  
M13, a cluster with an horizontal branch very similar to NGC6752.
Our simulations for NGC6752 horizontal branch predict however the presence of 
a significant fraction - at the level of $\sim50$\% - second generation stars along the cluster asymptotic giant branch. 
We conclude that there is no 
simple explanation for the lack of second generation stars in the spectroscopically surveyed sample, although the interplay 
between mass loss (with low rates) and radiative levitation may play a role in explaining this puzzle.
}
\keywords{stars: abundances -- stars: horizontal-branch -- stars: AGB and post-AGB -- 
globular clusters: general -- globular clusters: individual: NGC6752}
\authorrunning{Cassisi, S. et al.}
\titlerunning{Second generation AGB stars in NGC~6752}
  \maketitle


\section{Introduction}

In the last decade, the \lq{classical}\rq view of Galactic globular clusters (GGCs) as simple 
(single-age, single initial chemical composition) 
stellar populations has changed, thanks to a large body 
of photometric and spectroscopic observations \citep[see, e.g.,][and references therein for reviews]{gratton:12, piotto:12}. 
We now know that probably all GGCs host multiple stellar populations, each one with its own chemical abundance pattern. 
A typical GGC is populated by a stellar component born with the standard $\alpha$-enhanced metal distribution typical 
of the halo field population, plus additional sub-populations displaying 
light element (anti-)correlations, that is,  
a range of C and O (sometimes also Mg) depletions, together with  N and Na (sometimes also Al) enhancements.  
These additional sub-populations are characterized also by moderate (sometime large) helium enhancements \citep[see, i.e.,][]{milone:14}. 
Depending on the photometric filters \citep[see, i.e.,][]{sbordone:11, cassisi:13},  the 
cluster sub-populations may lie onto separate sequences in the 
colour-magnitude diagrams (CMDs), due to the effect of the multiple abundance distributions 
on isochrones and/or bolometric corrections.
This broad picture is nowadays denoted as the globular cluster \lq{multiple population phenomenon}\rq. 

The main scenarios devised to explain the formation of sub-populations 
within individual clusters  
\citep[see, e.g.][]{decressin, demink, dercole:11,conroy:11, valcarce:11} 
assume that the chemical patterns are produced by multiple star formation episodes 
during the early stages of the cluster evolution. After the formation
of a first generation (FG)  of stars 
with metal abundance ratios (and He abundance) typical of the halo field population, 
successive generations (SG stars) originate from 
matter ejected by preexisting FG stars ({\sl polluters}) -- 
massive asymptotic giant branch stars or massive stars --  
diluted with material of FG composition not yet involved in star formation episodes. 
A different scenario proposed recently by \citet{bastian:13} assumes instead that stars with SG composition are 
actually FG low mass stars polluted during their fully convective 
pre-main sequence phase (pre-MS -- hence their chemical composition at the
beginning of the MS is essentially uniform, like the multiple star
formation scenario) by the ejecta of short-lived massive stars. 
Throughout the paper we denote as FG stars the cluster sub-population 
with metal mixture and He abundances typical of the Galactic halo field population, and as SG stars the other populations, 
irrespective of their origin.

Tracing the distribution of SG 
stars along the observed cluster CMDs is very important to both study the evolutionary properties of SG stars, and  
shed more light on their formation and the nature of the {\sl polluters}.

One striking feature of SG stars is, at least in some clusters, 
the difference between the number ratios of SG to FG objects 
along the asymptotic giant branch (AGB) and the red giant branch (RGB).  
While stars with very strong CN bands --a signature, together with high-Na and low-O abundances, of SG 
composition-- are 
abundant along the RGB of any GGC, the majority of AGB stars in these
clusters usually seem to be FG objects. This early discovery by \cite{norris:81} has been confirmed 
by more recent investigations \citep{campbell:10, gdbcl:10}.
Using spectroscopic measurements of Na,   
\cite{campbell:13} have found no SG stars in a sample of AGB stars hosted by NGC6752. 
On the other hand, in clusters such as M~5 \citep{sn:93} and possibly 47~Tuc \citep{campbell:06}, CN-strong/Na-rich 
AGB stars have been observed, and the discrepancy with 
the ratios observed along the RGB is probably not large. 
It is evident that additional observational effort has to be devoted to this issue.

Recently, an attempt to connect the lower incidence of SG stars along the AGB 
to the multiple population phenomenon 
has been made by \cite{gdbcl:10} \citep[but see also the pioneering works by][]{norris:81, norris}. 
These authors found a correlation between the relative frequency of AGB stars 
and the minimum mass of stars along the horizontal branch 
(HB) in a given cluster (that is a \lq{proxy}\rq\ of the maximum effective temperature along the HB, and hence of the HB morphology), with a further dependence on the cluster metallicity. 
Clusters with the lowest number ratio of AGB to HB 
stars (the so called ${\rm R_2 }$ parameter, \citet{caputo:89}) are those with the bluest HB morphology. 
In fact, objects with extended HB blue tails like NGC~2808 and NGC~2419 display the lowest ${\rm R_2 }$ values, e.g.  
${\rm R_2\sim 0.06}$, to be compared with typical values ${\rm R_2}\sim$0.12 for redder morphologies \citep[see e.g.][]{sq00}, 
and the very high ${\rm R_2}=0.18\pm0.02$ for M5 \citep{sb:04}.

To connect this result to the cluster multiple populations, 
we recall that moving towards the blue side of the HB
(e.g. towards lower masses) when the evolving mass drops below $\sim$0.6${\rm M_{\odot}}$ (the exact value depending on the 
initial chemical composition)  
the ratio of HB to AGB timescales increases steadily, until reaching    
a minimum HB mass (equal to about 0.5 ${\rm M_{\odot}}$) below which stars do not evolve to the AGB phase 
-- the so-called {\sl AGB-manqu\'e} objects -- \citep{gr:90, dorman:93}. 
After central-He exhaustion these stars move to the white dwarf cooling sequence. 
SG stars are indeed expected to have a lower mass along the HB due to the fact that 
they are typically He-enhanced, hence they are originated by RGB stars with a lower mass (at a given age, 
He-enhanced stars have shorter main sequence lifetimes) and end up with a --on average-- lower mass along the HB 
if the RGB mass loss is approximately the same for FG and SG sub-populations 
\citep[see, e.g.][and references therein for a more detailed explanations]
{dantona:02, dalessandro:11, dalessandro:13, gratton:11, marino:11, marino:14}.
One expects therefore that the AGB of clusters with a blue HB should lack at least part of the SG component, 
compared to what is seen along the previous RGB phase. 

In this paper we have studied in detail the case of the AGB population of NGC6752, in light of 
the recent results by \cite{campbell:13}. 
These authors have performed an accurate spectroscopic analysis of both RGB and AGB stars, determining 
Na abundances. As also stated by these authors, Na abundance estimates in cool giants should be more robust than 
other light elements such as C and N, because they are less affected by molecular band formation and surface gravity estimates, 
and/or by the possibility of {\sl in situ} 
variations due to evolutionary process(es) \citep[see, e.g.][and references therein]{norris:81, scw, denissenkov:14}. 
\cite{campbell:13} found that {\sl all} 20 AGB stars in their sample display Na abundances 
typical of FG stars. To explain 
the lack of SG AGB stars and the observed low value of ${\rm R_2}$ (${\rm R_2 \sim 0.06}$), these authors
envisaged the following scenario. 
All FG stars populate the redder side of the observed HB, and  
all objects hotter than $\sim$11500~K --the 
so-called Grundahl's jump \citep{grundahl:99}, corresponding to the
observed effective temperature of the onset of radiative levitation for HB stars-- miss the AGB phase, even though standard 
stellar evolution predicts that at the cluster metallicity only stars evolving on the HB at ${\rm T_{eff}}$ larger 
than $\sim$24000~K do skip this phase \citep{dorman:93, basti1}.

The implication is that objects in the ${\rm T_{eff}}$ range between $\sim$11500~K and at least $\sim$24000~K, must have 
lost mass very efficiently during the HB phase, to avoid the AGB phase. 
We note that the effect of an efficient mass loss during the HB evolution has been   
already explored in previous analyses \citep[see, e.g.][and references therein]{yong:00}. By employing the 
Reimers \citep{reimers} mass loss law, \cite{campbell:13} estimated a value for the free parameter
 $\eta \sim$10 to explain the lack of SG 
stars along the AGB, an enhancement of a factor $\sim$20 compared to 
the standard value used in the calculation of RGB models. Such an enhancement compared to standard values of $\eta$ 
is tentatively attributed 
to the metal enhancement in the envelopes of these stars, caused by radiative levitation 
\citep[see, e.g.,][]{michaud:08}.
If the evolution of NGC6752 HB stars is \lq{typical}\rq\ of GGCs with 
long blue HB tails, this scenario implies that a cluster like M13, with a very similar HB morphology, [Fe/H] and age, 
should display the same lack of SG stars along the AGB.

In this analysis we have 
first revisited the HB mass loss scenario discussed by
\cite{campbell:13}, considering additional constraints from 
independent theoretical considerations and observations (Section~2). 
As a second step, we have performed synthetic HB modelling, using canonical HB models without 
mass loss during the core He-burning stage, 
to establish more accurately whether the observed low value of 
${\rm R_2\sim 0.06}$ does really require very efficient mass loss along the
HB. We have then studied the AGB population in our synthetic samples,
to check its expected composition in terms of FG and SG components (Section~3).   
We conclude with a discussion of the results and a comparison with the abundances 
measured in M13, a cluster whose properties are very similar to those of NGC~6752 (Section~4).

\section{Mass loss on the hot HB}
 
In the scenario envisaged by \cite{campbell:13}, all NGC6752 HB stars hotter than  ${\rm T_{eff}\sim11,500}$~K 
fail to reach the AGB due to enhanced mass loss, possibly associated to the surface metal enhancement caused 
by radiative levitation. This temperature is assumed to mark the transition between FG and SG stars along the cluster HB, 
consistent with spectroscopic measurements \citep{villanova:09} of nine HB stars in the   
${\rm T_{eff}}$ range between $\sim 8500$~K and $\sim 8800$~K, that show how 
all objects but one have [Na/Fe] ratios typical of FG stars, and the average He abundance is consistent also 
with the FG component.

\begin{figure}
\centering
\includegraphics[scale=.4500]{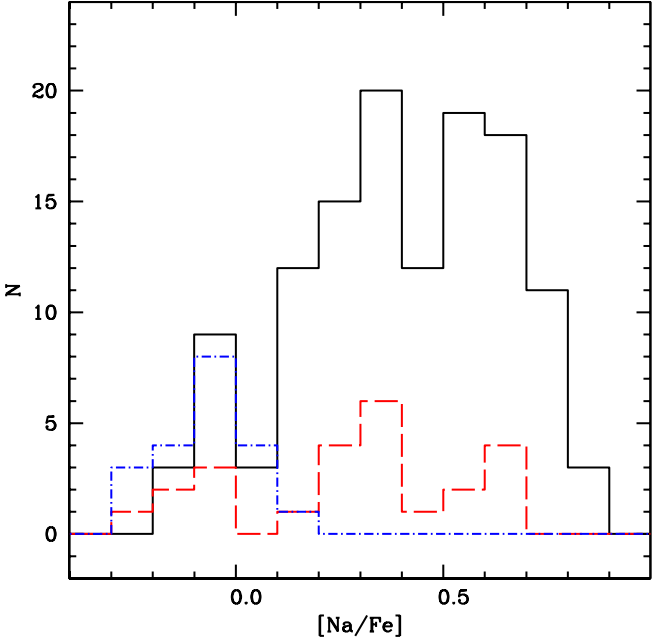}
\caption{Observed distributions of [Na/Fe] abundance ratios in NGC6752 RGB stars as determined by \cite{campbell:13} 
and \cite{cbg:07} --dashed and solid line, respectively-- and in AGB stars 
\citep[][dash-dotted line --see text for details]{campbell:13}.
}
\label{abundances}
\end{figure}

As discussed in the introduction, this enhanced mass loss efficiency is invoked by \cite{campbell:13} to explain both 
the observed low value of ${\rm R_2}$,   
and the observed [Na/Fe] distribution along the AGB when compared to the RGB counterpart, as displayed by Fig.~\ref{abundances}.
In the same figure we have shown also the [Na/Fe] RGB distribution determined by \cite{cbg:07} with a larger sample of stars.
A Kolmogorov-Smirnov test shows that the two RGB abundance distributions determined by \cite{campbell:13} and \cite{cbg:07}  
are statistically consistent.
The three \rq{peaks}\lq\ in the histogram of the RGB abundances correspond to [Na/Fe]$\sim -$0.05 (FG), $\sim$+0.35 and 
$\sim$+0.60 (both SG), respectively, and can be nicely associated on a one-to-one basis to the components 
of the cluster triple MS discovered by 
\cite{mmp:13}. 
It is very clear that the AGB abundances match the distribution of the lowest peak of the RGB Na distribution, corresponding 
to the FG composition. 

\begin{figure*}
\centering
\vskip -4.5cm
\includegraphics[scale=.850]{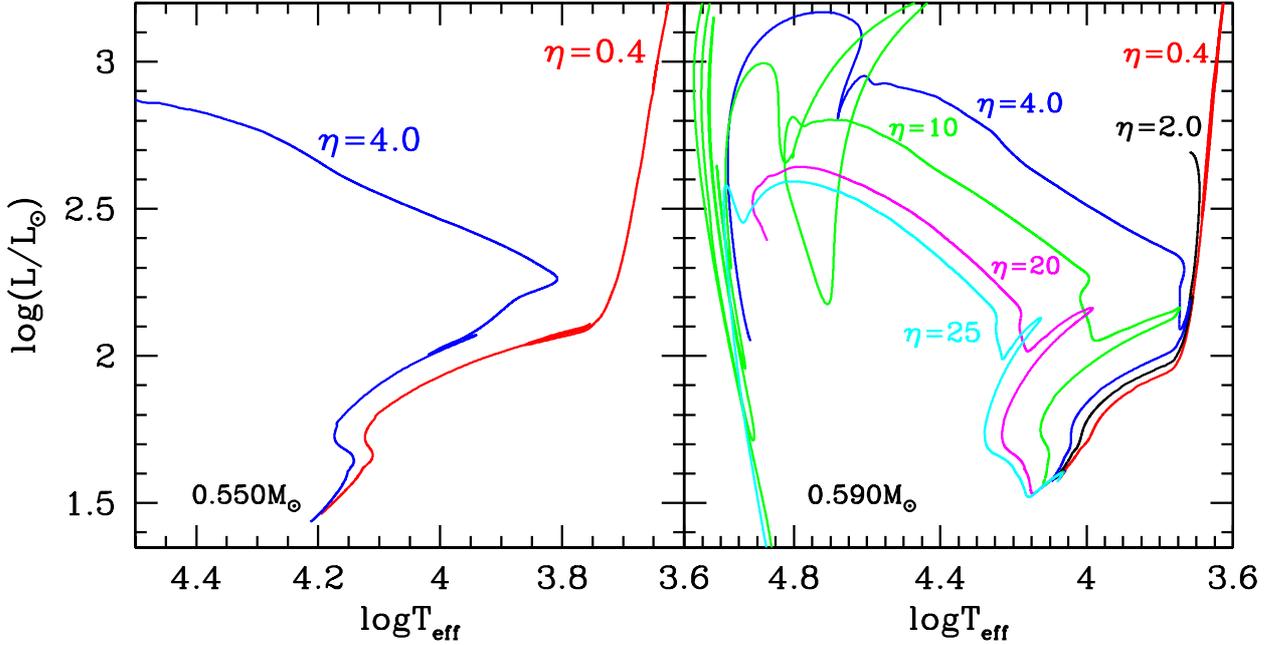}
\vskip -4.0cm
\caption{{\sl Right panel}: Hertzsprung-Russell (H-R) 
diagrams of stellar models with initial mass equal to ${\rm 0.59M_\odot}$, evolved from the ZAHB  
using the Reimers' mass loss law with various 
values of the free parameter $\eta$ (see labels). {\sl Left panel}: as
the right panel but for a model with initial mass equal to ${\rm 0.55M_\odot}$.}
\label{hr1}
\end{figure*}

Regarding the overall fraction and He abundance of SG stars in the cluster,  
the photometric analysis by \cite{mmp:13} revealed the presence of a triple MS with 
a \lq{normal}\rq\ cosmological Y$\sim$0.245 (the FG component),  
Y$\sim$0.254 and $\sim$0.27-0.28 respectively, these
latter two sequences being identified as the SG component. The number ratios among the three populations, that could be traced 
also along sub-giant branch and RGB  
are approximately 30:40:30, that correspond to a  
$\sim$70\% fraction of SG stars. This fraction is consistent with the RGB distribution of Na,  
displayed in Fig.~\ref{abundances}. \cite{mmp:13} do not find any evidence 
radial gradients of the population ratios out to about 2.5 half-mass radii.

Our first test was to revisit \cite{campbell:13} results about the required efficiency of 
the Reimers mass loss law to deplete the AGB population of NGC6752. To this purpose, we have calculated $\alpha$-enhanced 
([$\alpha$/Fe]=0.4) 
HB models with Y=0.273, Z=0.001 --corresponding to [Fe/H]$\sim -$1.6, consistent with the spectroscopic estimate 
[Fe/H]=$-$1.56 reported by \cite{cbg10}, and He-abundance 
appropriate for the more He-rich SG population of NGC6752-- for a large range of masses, with the 
same code, input physics, metal mixture and procedures of the BaSTI database \citep[][]{basti2}
\footnote{The whole BaSTI library is available at the URL: http://www.oa-teramo.inaf.it/BASTI.}.
From the BaSTI database we have considered a set of $\alpha-$enhanced HB stellar models for
Y=0.246, Z=0.001, the counterpart of the FG cluster composition, that 
will be used later on in the next section\footnote{Although SG stars appear to be O-depleted compared to the 
FG counterpart, as long as the ${\rm C+N+O}$ sum is kept constant, the evolutionary and structural properties of 
HB stars are barely affected \citep{basti:09}. Given that 
SG stars in NGC~6752 do not show any evidence of CNO enhancement \citep{cgl05, ymg13}, our assumption of adopting the same $\alpha-$enhanced mixture
for stellar models representing both FG and SG stars is justified.}

From the calculations with Y=0.273 we singled out the track with mass equal to ${\rm 0.59M_\odot}$, whose Zero Age HB (ZAHB) ${\rm T_{eff}}$ 
is $\sim$ 11500~K --corresponding to the Grundahl's jump-- and a hotter one of ${\rm 0.55M_\odot}$, 
with ZAHB ${\rm T_{eff}}\sim$ 16000~K. 
For these two masses we computed additional models considering the Reimers mass loss law with several choices of the 
free parameter $\eta$. 
The Hertzsprung-Russell (H-R) diagrams of the resulting evolutionary tracks are displayed in Fig.~\ref{hr1} 

The ${\rm 0.59M_\odot}$ tracks reach the AGB for $\eta$ increasing up to $\sim$2.0. This latter model 
climbs the AGB to a luminosity ${\rm log(L/L_\odot)\approx2.7}$, before moving towards the
hot side of the H-R diagram. The track with 
$\eta=4$ barely reaches the AGB stage, whereas for $\eta \ge 10.0$ the
track behaves as an AGB-manqu{\'e} star. 
Typical mass loss rates for the tracks with $\eta=10$ are a few times $10^{-9} {\rm M_{\odot} yr^{-1}}$.

As for the hotter ${\rm 0.55M_\odot}$ models, the tracks become AGB-manqu{\'e} for $\eta \ge 4$. Also in this case 
the critical mass loss rates are a few times $10^{-9} {\rm M_{\odot} yr^{-1}}$.
These \lq{threshold}\rq\ values of $\eta$ are a factor 10-25 higher than the typical values adopted for 
RGB evolutionary calculations with 
mass loss \citep[see i.e.][]{basti1, basti2, campbell:13}. In fact, employing such 
high values of $\eta$ would prevent RGB models from experiencing the He-flash \citep{cc:93}.

Our results are qualitatively 
in agreement with those obtained by \cite{campbell:13}, even though these latter authors 
considered a more massive HB model with a higher Y. 
These mass loss rates are
however much larger than constraints from the more physically motivated
mass loss law by   
\cite{vc:02}. These authors computed HB mass-loss rates under the hypothesis that radiation 
pressure on spectral lines drives 
a stellar wind during the He-burning stage; their mass loss
law\footnote{${\rm log~\dot{M}_{VC02} }= -11.70+1.07~{\rm log} ({\rm T_{eff}/20000)} 
                  +2.13~({\rm log(L/L_\odot) - 1.5})-1.09~{\rm log}({\rm 2M/M_\odot}) 
                  +0.97~{\rm log(Z/Z_\odot)}$, where ${\rm
    T_{eff}}$ is in Kelvin, and the other symbols have their usual
  meaning. } is expected to be much more appropriate 
for HB models than the Reimers formula, and is applicable in the range
between 12500 and 35000~K.  The predicted rates are $\sim$2-3 orders of magnitude
lower than required. Even considering solar surface abundances for
the HB models, to mimic the effect of radiative
levitation\footnote{Spectroscopic analyses \citep[see e.g.][]{behr:03, pace:06} have shown that iron and
other heavy elements such as Ti and Cr can reach values around solar
or even higher, as a consequence of radiative levitation in the
atmosphere}  we found that the rates predicted by   \cite{vc:02}
equation must be increased by a factor $\sim$20-50 (depending on the mass of the HB
model) in order to attain the mass loss efficiency required to force a HB star to miss the AGB stage.

There are also some independent constraints on the efficiency of mass
loss in HB stars, albeit for very hot objects, in a ${\rm
  T_{eff}}$ range where models at constant mass already skip the AGB.
The constraints arise from the fact that mass loss affects the efficiency of 
atomic diffusion and radiative levitation, and mass loss rates too high  
would produce surface abundances inconsistent with observations.
\cite{michaud:11} considered a HB 
model with ${\rm T_{eff}\approx30000}$~K and determined that mass loss
rates compatible with the observed abundances for these stars can not be larger than
${\rm (3-5)\times10^{-14}}{\rm M_{\odot} yr^{-1}}$.
Another independent constraint on the mass loss efficiency in hot HB stars 
come from the evidence that some sdB stars undergo  
non radial pulsations. A recent analysis \citep[see][and references therein]{hu:11} shows that,
if mass loss is efficient in these objects, the rates cannot by larger than ${\rm \sim
  10^{-15}M_\odot/yr}$, otherwise the driving mechanism of the
pulsations  (an opacity-driven mechanism enabled by Fe that accumulate 
diffusively in the stellar envelope) would not work.
Assuming the mass loss scaling (rather than the absolute values) along the HB 
predicted by \cite{vc:02} is correct, the expected variation between temperatures of 
the order of 30000~K and $\sim$12000~K is within a factor 2-3. 
In conclusion, the huge mass loss rates required to force {\sl all} HB stars
with ZAHB location hotter than ${\rm T_{eff}\approx11500}$~K to miss the AGB stage seem to be 
disfavoured by the previous theoretical and observational considerations.
In any case, additional work on both the theoretical and observational side has to be done in order to 
settle this issue.

\section{Synthetic HB simulations}

\begin{figure}
\centering
\includegraphics[scale=.4500]{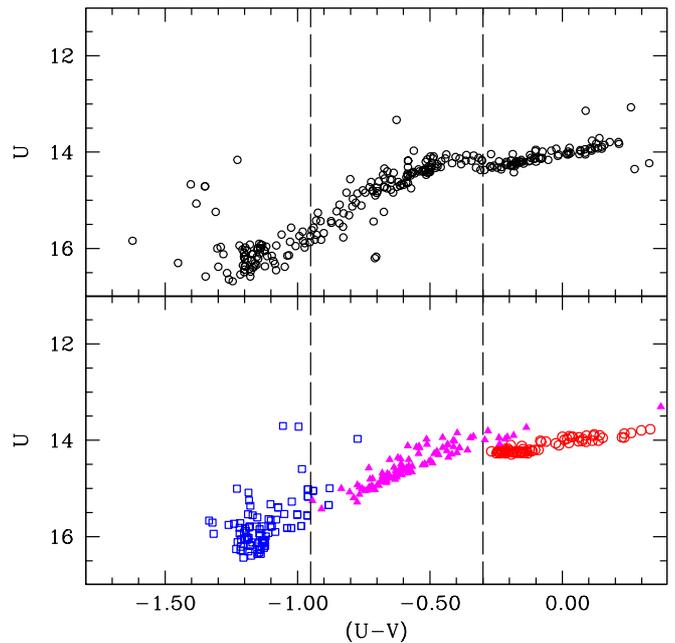}
\caption{{\sl Upper panel:} CMD of NGC~6752 HB stars from \cite{momany:02}. {\sl Lower panel:} One synthetic 
realization of the cluster HB, with the same number of stars as
observed. Open circles denote FG stars, filled triangles SG stars with
Y=0.254, open squares SG stars with Y=0.273, respectively. The vertical lines define the boundaries of the three boxes 
in which the observed CMD has been partitioned. The location of
the reddest vertical line coincides with the location of the Grundahl's jump (see text for details).
}
\label{hbsynt1}
\end{figure}

With the goal to determine whether the observed value of ${\rm R_2}$  is consistent with a large 
fraction of HB stars missing the AGB, we have performed an analysis based
on synthetic HB calculations \citep[see, e.g.,][for details about this technique]{r1973, catelan}, made with the code described by 
\cite{dalessandro:13}. As observational counterpart 
we have considered the UV CMD by \cite{momany:02}, whose HB (cleared of the field contamination) is displayed in 
Fig.~\ref{hbsynt1}. 
The shape of the cluster blue HB in the U-(U-V) CMD is more sensitive to the He-abundance than 
\lq{classical}\rq optical CMDs.
As for the theoretical models, we have employed 
BaSTI $\alpha$-enhanced HB tracks for Z=0.001, with Y=0.246 and Y=0.273, described in the previous section. We are assuming that the 
age of SG and FG stars is the same, equal to $12.5$~Gyr. This assumption
is appropriate because studies of multiple populations in GGCs have shown that age differences between FG and SG stars
is at most of the order of $\sim$100~Myr  \citep[see, e.g.][and references therein]{cassisi:08, marino:12}. The precise value of this  
common age is also not crucial, because a different absolute age would simply cause a variation of the RGB mass loss 
required to reproduce the observed mass distribution along the HB.
The input physics of BaSTI models is detailed in \cite{basti1} and references therein. Crucial to the prediction 
of HB and AGB evolutionary timescales is the treatment of core mixing during central He-burning. Our models 
include semiconvection with breathing pulses suppressed \citep[see][and references therein]{basti1}. 
Notice how also very recently \citet{gabriel} have shown that the inclusion of semiconvection 
in the calculation of HB models (that extends the He-burning lifetime and the final He-depleted core mass) 
seems to be an adequate approach to treat core mixing during central He-burning.
As for the breathing pulses (a mixing instability occurring at the boundary of the 
convective core when central He is substantially depleted), the analysis by \citet{cassisi:01} shows 
that their inclusion would smooth out the AGB clump, at odds with the observations, that display very 
concentrated and prominent AGB clumps in well populated photometries of GGCs \citep[see, e.g., M3 and 47Tuc   
photometries by][respectively, as just two examples]{buonanno, beccari}. As discussed by \citet{cassisi:03},  
the different techniques available to suppress this instability do not affect the predicted value of ${\rm R_2}$.
In general, the inclusion of breathing pulses would decrease the predicted value 
of ${\rm R_2}$ by $\sim$30\% for models at the red side of the HB.

Once the mixing treatment is fixed, the main source of uncertainty in the predicted ${\rm R_2}$ values is given 
by the current uncertainty in the ${\rm ^{12}C(\alpha,\gamma)^{16}O}$, that affect this parameter at the level of 
$\pm$0.01.

Detailed analyses of the uncertainties affecting He-burning, low-mass stellar models can be 
found in \cite{cassisi:98, cassisi:99, cassisi:01, cassisi:03, cassisi:07} and reference therein.


As described before, we considered HB models with two values of Y; synthetic stars with 
intermediate values of the initial He mass fraction were obtained by interpolation during the synthetic HB calculations. 
The BaSTI HB tracks include also the AGB evolution until the first thermal pulse (TP); given that 
in low mass stars the TP lifetime is largely 
negligible compared to the AGB timescales before the first TP 
\citep[see e.g. the recent AGB calculations by][for the case of 1.0${\rm M_{\odot}}$ models at various metallicities]{wf:09}, 
our simulations can predict consistently the value of ${\rm R_2}$ for a given HB morphology.
SG stars with enhanced He are represented by models with standard FG chemical composition and 
the appropriate He abundance.

To account for the effect of radiative levitation in the stellar atmospheres, that affects stars 
hotter than the Grundahl's jump, we have applied to the bolometric luminosities predicted by the models,  
bolometric corrections to the U and V photometric bands for 
[Fe/H]=0.0 (and scaled-solar mixture) when ${\rm T_{eff}}$ is above
$\sim$12000~K\footnote{Only the morphology of the evolutionary tracks is affected by the choice of 
the bolometric corrections,  
as shown in Fig.~7 of \cite{dalessandro:11}, not the evolutionary timescales.}. 
This is a crude approximation, due to the lack of extended grids
of both stellar evolution and atmosphere models with
a large range of chemical compositions, that include consistently 
the effect of radiative levitation during the whole HB evolution. However, 
the use of this approximation in synthetic HB modelling has allowed \cite{dalessandro:11} 
to recover the different magnitude levels of NGC~2808 sub-populations 
with the three He-abundances inferred from the multimodal main sequence, 
in optical-ultraviolet CMDs of the HB. Also, the ${\rm T_{eff}}$ 
distribution of NGC~2808 extreme HB stars determined from the simulations 
was found in good agreement with the spectroscopic estimates by \cite{Moh04}
As shown by Fig.~\ref{hbsynt1}, we divided the observed CMD of the cluster HB into three boxes, 
with $-0.3 < (U-V) \le 0.4$ (red box) , $-0.95 < (U-V) \le -0.3$ (intermediate box), 
and $(U-V) \le -0.95$ (blue box). The blue edge of the red box corresponds approximately 
to the colour of the Grundahl's jump, while the blue edge of the intermediate box  
is chosen to have approximately 30:40:30 number ratios of HB stars in the three 
regions (red:intermediate:blue), the same as the estimated ratios of the three cluster sub-populations  
(in order of increasing Y) along MS and RGB, that we use as input.

\begin{figure}
\centering
\includegraphics[scale=.4500]{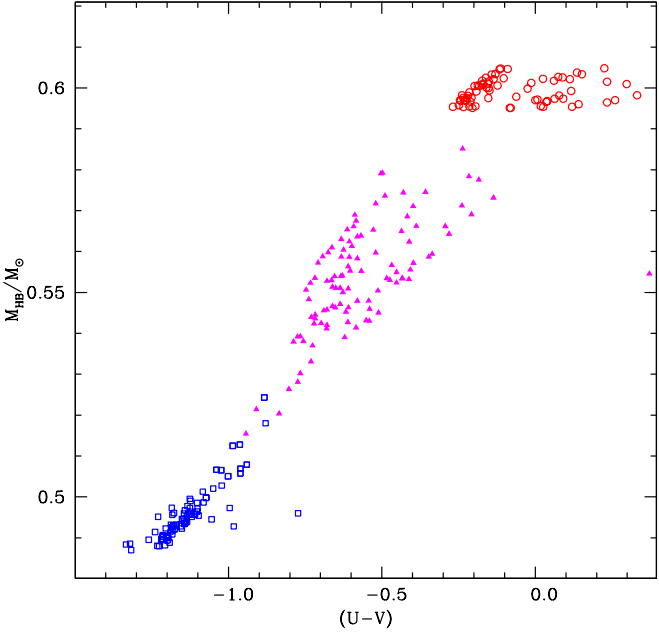}
\caption{Mass distribution along the synthetic HB of
  Fig.~\ref{hbsynt1}. The meaning of the symbols is as in Fig.~\ref{hbsynt1}}
\label{hbsynt1b}
\end{figure}

In our modelling we assumed that stars belonging to 
the three sub-populations are fed at a continuous rate onto the HB \citep[see, e.g.,][]{r1973}, and 
that the FG component is located in the red box, consistent with spectroscopic observations. 
The slope of the observed HB in the intermediate box is sensitive to Y, and is consistent 
with the location of synthetic stars from the SG component with Y=0.254.  
As a consequence, the most He-enriched stars must be largely confined to the blue box.
With these constraints on the simulations, the relative number of stars in the different boxes, plus the 
magnitude and colour location 
of the HB were reproduced with E(B-V)=0.03 \citep[consistent with E(B-V)=0.04$\pm$0.005 
and systematic error of $\pm$0.011, as estimated by][]{gbc03}, ${\rm (m-M)_0}$=13.12 \citep[consistent, i.e.,   
with the cluster white dwarf distance by][]{rbf96} 
and the mass distribution displayed in Fig.~\ref{hbsynt1b}. The simulations have included also a Gaussian 
random photometric error 
whose mean value for the various filters was taken from \cite{momany:02} photometry.

Figure~\ref{hbsynt1} shows one realization of the HB, with the number of synthetic stars (275) 
equal to the one in \cite{momany:02} photometry. As discussed before, empirical and theoretical 
constraints dictate that the three components should mainly populate separate regions, and in this 
synthetic CMD only evolved SG stars cross the boundaries between boxes, 
causing just a minor \lq{contamination}\rq\ of the neighbouring population. 

Assuming a common age of 12.5~Gyr, the mass distribution of Fig.~\ref{hbsynt1b} 
corresponds to the following total RGB mass loss (${\rm \Delta M}$) for the 
FG and SG populations:

\begin{enumerate}

\item{FG -- ${\rm \Delta M}$ ranging from 0.180 to 0.190~${\rm M_{\odot}}$ with a flat probability distribution;} 

\item{SG, Y=0.254 -- ${\rm \Delta M}$=(0.220 $\pm$ 0.014)~${\rm M_{\odot}}$ with a Gaussian probability distribution;}

\item{SG, Y=0.273 -- ${\rm \Delta M}$=(0.282$\pm$0.002)~${\rm M_{\odot}}$ Gaussian for 80\% of stars, and ${\rm \Delta M}$-
0.268$\pm$0.006 for 20\% of stars.}  

\end{enumerate}

Small variations of this age and/or possible small age differences between FG and SG stars will affect the 
${\rm \Delta M}$ values necessary to reproduce the mass distribution of Fig.~\ref{hbsynt1b}. For example, an 
age change by 0.5~Gyr 
around 12.5~Gyr changes the evolving RGB mass by $\sim$0.01~${\rm M_{\odot}}$.  

It is obvious from Fig.~\ref{hbsynt1} that an exact match of the observed colour (hence mass) distribution 
of HB stars within each box would require (especially in the intermediate and blue box) more complex mass loss laws. 
There is no \lq{a priori}\rq\ reason why a simple flat or Gaussian probability distribution should represent the 
correct mathematical form of the total RGB mass loss, but on the other hand they provide an approximation that is sufficient 
for most purposes. In fact, the value of the parameter ${\rm R_2}$ we obtained from our simulations is basically unaffected 
(changes by less than 0.01) by modifying the mass distributions within each box, i.e. varying slightly 
the mean value and/or the spread of the Gaussian distributions, or switching from Gaussian to 
flat distribution, while keeping unchanged the number of objects within each box.

Simulations with a large total number of HB stars 
(about 50 times the observed number in our adopted photometry) provide 
${\rm R_2}$=0.07, in very good agreement with the observed ${\rm R_2}$=0.06$\pm$0.02 \citep[see data in Table~1 of][]{sq00}.
When considering multiple realizations of the HB with the same number of stars in \cite{momany:02}, we obtain 
statistical variations of ${\rm R_2}$ --due to small number statistics-- of about $\pm$0.01 around ${\rm R_2}$=0.07.
Coincidentally, for the number of HB stars in our adopted CMD, one expects a number of AGB stars approximately equal 
to the number sampled spectroscopically by \cite{campbell:13}.

\begin{figure}
\centering
\includegraphics[scale=.480]{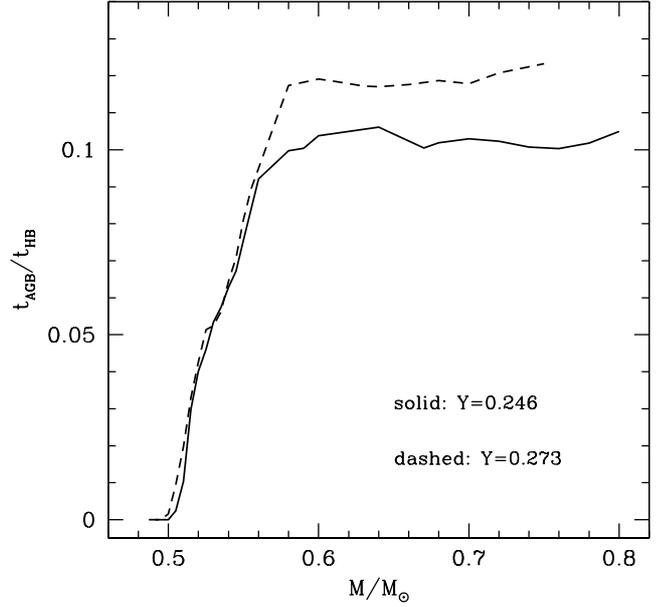}
\caption{Ratio of AGB to HB lifetimes as a function of the total mass,  
for the models used in our simulations (see text for details)}
\label{ratios}
\end{figure}

This agreement with our models at constant mass disfavours the possibility that a large number 
of stars miss the AGB phase due to mass loss during the core He-burning stage, 
apart from the bluest object that \lq{naturally}\rq\ skip the AGB. 
In the hypothesis that all SG stars along the HB miss the AGB phase, our simulations 
predict a too small value ${\rm R_2}$=0.03.
The agreement with the observed value of ${\rm R_2}$ comes instead from the decreasing ratio between AGB and HB lifetimes 
when moving towards the blue side of the HB. This is shown clearly by Fig.~\ref{ratios}, that 
displays the ratio of AGB to HB lifetimes for the models used in our simulations. 
Notice the steep drop below $\sim$0.56-0.58${\rm M_{\odot}}$ --for the sake of comparison 
the upper mass limit in the synthetic CMD is $\sim$0.6${\rm M_{\odot}}$, see Fig.~\ref{hbsynt1b}. 
In this region the lifetime ratio is independent of the composition. 
He-rich models have slightly higher ratios above this mass threshold, because of a more efficient H-burning shell 
during HB phase.  

As a test, we have calculated a well populated synthetic HB with the parameters listed above, 
considering just the FG component. 
We obtained in this case ${\rm R_2}$=0.11. The same test considering only the SG population with Y=0.254 provided ${\rm R_2}$=0.09, 
while the case with 
just the SG population with Y=0.273 gave ${\rm R_2}$=0.02. The progression towards decreasing values of ${\rm R_2}$ when 
considering increasingly bluer 
morphologies is clear. 
It is important to notice that all stars belonging to the SG
population with Y=0.254 are expected to evolve to the  AGB. All stars predicted to 
miss the AGB belong to the SG population 
with Y=0.273; however, also a very small fraction of this population is expected
to evolve to the AGB. 

By examining the AGB population in our simulations we found that
about 50\% of the stars belong to the FG, and about 50\% to the SG
sub-population with Y=0.254.
From multiple synthetic realizations of the HB we determined a
probability of only  $\sim$60\% to find a single AGB star belonging to the
SG with Y=0.273 within a sample of 20 AGB objects, i.e. the size of
\cite{campbell:13} AGB sample.

We close this section by recalling that also the recent study by \citet{ccd} considered 
the enhanced initial He in SG stars as the likely explanation for the results by \citet{campbell:13}. 
The authors however did not perform any detailed synthetic HB simulations  
and employed a somewhat larger He-abundance range in SG stars compared to the current 
observational constraints (see also their discussion 
on this subject).

\subsection{Consistency checks}

To establish whether our theoretical predictions of $R_2 $ are biased by possible 
systematic errors  
in the evolutionary lifetimes predicted by the HB tracks hence the agreement 
with NGC6752 is just fortuitous, we have performed additional tests.
The most important one was to check whether our simulations are able to reproduce the value of ${\rm R_2}$ also for a 
cluster with a redder morphology, for which one expects all HB stars to end up evolving along the AGB. 
To this purpose we examined the detailed
simulations of the HB of M3 --a cluster with metallicity very similar
to NGC~6752 but with essentially all HB stars cooler than the Grundahl's jump--  
presented in \cite{dalessandro:13}, 
obtained with the same evolutionary tracks and code employed here. 
The work by \cite{dalessandro:13} makes use of 
far-ultraviolet(UV)/optical colour-magnitude diagrams for clusters
with extended HBs, that enable to
estimate the initial He abundance along the HB blue-tails. This is because 
in UV filters the HB becomes almost horizontal at high effective
temperatures, and He variations cause variations of the UV magnitude
levels \citep[see][for details]{dalessandro:11, dalessandro:13}.

The analysis by \cite{cho} determined an observed value
${\rm R_2}$=0.12$\pm$0.03 for M3 
(we estimated the error bar on ${\rm R_2}$ using Poisson statistics). 
The simulations by \cite{dalessandro:13} match the observed stellar
distribution and magnitude levels along the cluster HB with 
a range of initial Y abundances equal to $\sim$0.02, starting from a
minimum Y=0.246. From these simulations we derived  
${\rm R_2}$=0.11, in agreement with the observed value.
 
We have then considered another cluster with an extended blue HB 
and basically the same metallicity and age as NGC6752, 
namely M13 \citep[see e.g.][and references therein]{dalessandro:13}. 
The HB ratio, ${\rm HB_R = (B-R)/(B+V+R)}$ --where B, V and R are 
the number of stars at the blue side, within, and at the red side 
of the RR Lyrae instability strip, respectively-- 
is 0.97 for M 13 and 1.0 for NGC 6752 \citep[see e.g.][]{cbg:07}.

The study by \cite{cho} derived  
${\rm R_2=0.07\pm0.03}$, consistent with the result for NGC~6752. 
Our detailed simulations for this cluster HB have been discussed by 
\cite{dalessandro:13}, who estimated a range
of initial Y from 0.246 up to $\sim$ 0.31. This inferred higher range of Y values compared 
to NGC6752 is consistent with the fact that M13 displays a larger range of Na abundances. 
The simulations provide obtained ${\rm R_2}$=0.06 (as for NGC6752) again in very good
agreement with the observations.

The study by \cite{cho} provides also another very interesting test for the evolutionary timescales 
of the models. These authors determined ${\rm R_2=0.16\pm0.07}$ (we calculated the error bar from Poisson statistics) 
when considering all AGB stars but only the reddest component along the HB -- e.g. 
all HB stars  
located at ${\rm T_{eff}}$ roughly lower than the Grundahl's jump location. 
On the simulation side, this reddest HB component corresponds 
to the synthetic HB stars with Y below 0.26 in the analysis by \cite{dalessandro:13}. 
Restricting the synthetic HB population to this component, while keeping the full AGB 
population, the simulations provide ${\rm R_2}$=0.15, again in agreement
with observations.
It is worth noticing that, within this test, if synthetic stars with Y higher than 0.26 (all HB stars 
hotter than the Grundahl's jump) do not contribute to the AGB population, the 
simulations would give ${\rm R_2}$=0.11, much lower than the
value observed. Unfortunately the large statistical error on the 
observed ${\rm R_2}$ of this test provides only 
an additional qualitative indication that HB models at this metallicity
predict ${\rm R_2}$ values consistent with observations.

\section{Discussion and conclusions}

In this paper we have explored the scenario proposed by 
\cite{campbell:13} to explain the [Na/Fe] abundances measured along the AGB of NGC6752, and the low 
value of the ${\rm R_2}$ parameter, that envisages strong mass loss for all SG HB stars (located at ${\rm T_{eff}}$ 
hotter than the temperature for the onset of radiative levitation), that prevents 
their evolution to the AGB.
We have found that the required mass loss rates are of the order of 
$10^{-9} {\rm M_{\odot} yr^{-1}}$, much higher than 
current empirical and theoretical constraints. 

Our synthetic modelling --with our adopted stellar models-- of the observed HB has provided a 
value of ${\rm R_2}$ consistent with the observations. 
Imposing that SG stars do not reach the AGB phase would produce a value of 
${\rm R_2}$ too low compared to the observed one.
Consistent synthetic HB analyses of M3 (a cluster with the same metallicity, age but a much redder 
HB morphology) and M13 (a cluster with also almost the same HB morphology of NGC6752) provide 
${\rm R_2}$ values again in agreement with observations.
As a conclusion, our HB models predict 
correctly ${\rm R_2}$, at least at the metallicity of NGC6752, and no strong mass loss during the core He-burning stage is required 
to reproduce the observed cluster ${\rm R_2}$.

The HB simulations also predict that for the 
size of NGC6752 spectroscopic AGB sample, about 50\% of the population is made of FG stars, and the rest 
belongs to the SG population with the lower Y mass fraction.
The theoretical prediction that SG stars populate the AGB is consistent with the spectroscopic analysis by 
\cite{villanova:09}, who found in their small sample of HB stars cooler than the Grundahl's jump, 
one object with high Na, brighter than the bulk of the HB stars. This object can be understood as an evolved HB  
SG star, that is moving towards the AGB. The presence 
of some evolved SG stars at ${\rm T_{eff}}$ below the Grundahl's jump is  
expected from the HB simulations (see Fig.~\ref{hbsynt1}).

On the other hand, only FG stars are observed in the AGB sample analyzed spectroscopically (see Fig.~\ref{abundances}). 
This is hard to explain, given the indications presented in this paper 
against the hypothesised enhanced mass loss. 
The situation gets even more puzzling when considering the results for AGB stars in M13. The 
spectroscopic analysis of a sample of its AGB and RGB stars 
by \cite{jp:12} shows that the AGB stars span almost the whole 
range of [Na/Fe] values covered by RGB stars. There is no indication that only FG stars 
inhabit the AGB.
Given the close similarity of M13 and NGC6752 in terms also of the HB morphology, it is difficult to 
find a mechanism to deplete SG AGB stars that works in NGC6752 but not in M13.

We conclude mentioning a final point that may be relevant to this issue. 
\cite{campbell:13} measured Na abundances in RGB and AGB stars, and derived [Na/Fe] ratios 
by considering a uniform [Fe/H] value for all stars, as determined in the literature on RGB stars.
In case of M13, \cite{jp:12} determined both Na and Fe abundances for their AGB stars.
What would be important to consider is the effect of the interplay between radiative levitation and mass loss. 
For example, \cite{michaud:11} show that 
both Na and Fe are strongly enhanced in the outer layers of a 0.59 ${\rm M_{\odot}}$ (initial Z=0.0001) 
HB model at 14000~K. If mass loss is efficient even only at 
the low level required to  preserve the efficiency of radiative levitation, 
some amount of Na and Fe is certainly lost from the star. 
When convection homogenises the remaining 
envelope on the AGB, the surface abundances of these two elements will be depleted to some degree (maybe negligible) 
compared to the RGB abundances. 
If both Na and Fe are depleted by roughly the same amount, the measured [Na/Fe] on the AGB should reflect 
the original ratios along the RGB. If however only Na is measured and Fe is assumed equal to the RGB values, 
one may in principle underestimate the AGB [Na/Fe] ratio for objects affected by radiative levitation along the HB 
(as it is the case of SG stars in NGC6752 and M13).

At the moment there are no available calculations of the full evolution from the HB to the AGB, self-consistently accounting for
diffusion and radiative levitation, and mass loss. 
It would be certainly important to attempt these computations, with the aim of establishing whether 
the interplay between radiative levitation and mass loss plays a major role in explaining the puzzling 
chemical composition of AGB stars in NGC6752 and the difference with M13.

\begin{acknowledgements}
We warmly thank R. Gratton for the very useful suggestions and discussions about  
uncertainties of light element spectroscopy in globular cluster cool giants.
SC acknowledges financial support
from PRIN-INAF 2011 "Multiple Populations in Globular Clusters: their
role in the Galaxy assembly" (PI: E. Carretta), and from PRIN MIUR 2010-2011,
project \lq{The Chemical and Dynamical Evolution of the Milky Way and Local Group Galaxies}\rq, prot. 2010LY5N2T (PI: F. Matteucci). 
\end{acknowledgements}

\bibliographystyle{aa}
\bibliography{massloss_hbagb}

\begin{thebibliography}{70}
\expandafter\ifx\csname natexlab\endcsname\relax\def\natexlab#1{#1}\fi

\bibitem[{{Bastian} {et~al.}(2013){Bastian}, {Lamers}, {de Mink}, {Longmore},
  {Goodwin}, \& {Gieles}}]{bastian:13}
{Bastian}, N., {Lamers}, H.~J.~G.~L.~M., {de Mink}, S.~E., {et~al.} 2013,
  \mnras, 436, 2398

\bibitem[{{Beccari} {et~al.}(2006){Beccari}, {Ferraro}, {Lanzoni}, \&
  {Bellazzini}}]{beccari}
{Beccari}, G., {Ferraro}, F.~R., {Lanzoni}, B., \& {Bellazzini}, M. 2006,
  \apjl, 652, L121

\bibitem[{{Behr}(2003)}]{behr:03}
{Behr}, B.~B. 2003, \apjs, 149, 67

\bibitem[{{Buonanno} {et~al.}(1994){Buonanno}, {Corsi}, {Buzzoni}, {Cacciari},
  {Ferraro}, \& {Fusi Pecci}}]{buonanno}
{Buonanno}, R., {Corsi}, C.~E., {Buzzoni}, A., {et~al.} 1994, \aap, 290, 69

\bibitem[{{Campbell} {et~al.}(2013){Campbell}, {D'Orazi}, {Yong},
  {Constantino}, {Lattanzio}, {Stancliffe}, {Angelou}, {Wylie-de Boer}, \&
  {Grundahl}}]{campbell:13}
{Campbell}, S.~W., {D'Orazi}, V., {Yong}, D., {et~al.} 2013, \nat, 498, 198

\bibitem[{{Campbell} {et~al.}(2006){Campbell}, {Lattanzio}, \&
  {Elliott}}]{campbell:06}
{Campbell}, S.~W., {Lattanzio}, J.~C., \& {Elliott}, L.~M. 2006, \memsai, 77,
  864

\bibitem[{{Campbell} {et~al.}(2010){Campbell}, {Yong}, {Wylie-de Boer},
  {Stancliffe}, {Lattanzio}, {Angelou}, {Grundahl}, \& {Sneden}}]{campbell:10}
{Campbell}, S.~W., {Yong}, D., {Wylie-de Boer}, E.~C., {et~al.} 2010, \memsai,
  81, 1004

\bibitem[{{Caputo} {et~al.}(1989){Caputo}, {Tornambe}, \&
  {Castellani}}]{caputo:89}
{Caputo}, F., {Tornambe}, A., \& {Castellani}, V. 1989, \aap, 222, 121

\bibitem[{{Carretta} {et~al.}(2007){Carretta}, {Bragaglia}, {Gratton},
  {Lucatello}, \& {Momany}}]{cbg:07}
{Carretta}, E., {Bragaglia}, A., {Gratton}, R.~G., {Lucatello}, S., \&
  {Momany}, Y. 2007, \aap, 464, 927

\bibitem[{{Carretta} {et~al.}(2010){Carretta}, {Bragaglia}, {Gratton},
  {Recio-Blanco}, {Lucatello}, {D'Orazi}, \& {Cassisi}}]{cbg10}
{Carretta}, E., {Bragaglia}, A., {Gratton}, R.~G., {et~al.} 2010, \aap, 516,
  A55

\bibitem[{{Carretta} {et~al.}(2005){Carretta}, {Gratton}, {Lucatello},
  {Bragaglia}, \& {Bonifacio}}]{cgl05}
{Carretta}, E., {Gratton}, R.~G., {Lucatello}, S., {Bragaglia}, A., \&
  {Bonifacio}, P. 2005, \aap, 433, 597

\bibitem[{{Cassisi} {et~al.}(2001){Cassisi}, {Castellani}, {Degl'Innocenti},
  {Piotto}, \& {Salaris}}]{cassisi:01}
{Cassisi}, S., {Castellani}, V., {Degl'Innocenti}, S., {Piotto}, G., \&
  {Salaris}, M. 2001, \aap, 366, 578

\bibitem[{{Cassisi} {et~al.}(1999){Cassisi}, {Castellani}, {degl'Innocenti},
  {Salaris}, \& {Weiss}}]{cassisi:99}
{Cassisi}, S., {Castellani}, V., {degl'Innocenti}, S., {Salaris}, M., \&
  {Weiss}, A. 1999, \aaps, 134, 103

\bibitem[{{Cassisi} {et~al.}(1998){Cassisi}, {Castellani}, {degl'Innocenti}, \&
  {Weiss}}]{cassisi:98}
{Cassisi}, S., {Castellani}, V., {degl'Innocenti}, S., \& {Weiss}, A. 1998,
  \aaps, 129, 267

\bibitem[{{Cassisi} {et~al.}(2013){Cassisi}, {Mucciarelli}, {Pietrinferni},
  {Salaris}, \& {Ferguson}}]{cassisi:13}
{Cassisi}, S., {Mucciarelli}, A., {Pietrinferni}, A., {Salaris}, M., \&
  {Ferguson}, J. 2013, \aap, 554, A19

\bibitem[{{Cassisi} {et~al.}(2007){Cassisi}, {Potekhin}, {Pietrinferni},
  {Catelan}, \& {Salaris}}]{cassisi:07}
{Cassisi}, S., {Potekhin}, A.~Y., {Pietrinferni}, A., {Catelan}, M., \&
  {Salaris}, M. 2007, \apj, 661, 1094

\bibitem[{{Cassisi} {et~al.}(2003){Cassisi}, {Salaris}, \&
  {Irwin}}]{cassisi:03}
{Cassisi}, S., {Salaris}, M., \& {Irwin}, A.~W. 2003, \apj, 588, 862

\bibitem[{{Cassisi} {et~al.}(2008){Cassisi}, {Salaris}, {Pietrinferni},
  {Piotto}, {Milone}, {Bedin}, \& {Anderson}}]{cassisi:08}
{Cassisi}, S., {Salaris}, M., {Pietrinferni}, A., {et~al.} 2008, \apjl, 672,
  L115

\bibitem[{{Castellani} \& {Castellani}(1993)}]{cc:93}
{Castellani}, M. \& {Castellani}, V. 1993, \apj, 407, 649

\bibitem[{{Catelan}(1993)}]{catelan}
{Catelan}, M. 1993, \aaps, 98, 547

\bibitem[{{Charbonnel} {et~al.}(2013){Charbonnel}, {Chantereau}, {Decressin},
  {Meynet}, \& {Schaerer}}]{ccd}
{Charbonnel}, C., {Chantereau}, W., {Decressin}, T., {Meynet}, G., \&
  {Schaerer}, D. 2013, \aap, 557, L17

\bibitem[{{Cho} {et~al.}(2005){Cho}, {Lee}, {Jeon}, \& {Sim}}]{cho}
{Cho}, D.-H., {Lee}, S.-G., {Jeon}, Y.-B., \& {Sim}, K.~J. 2005, \aj, 129, 1922

\bibitem[{{Conroy} \& {Spergel}(2011)}]{conroy:11}
{Conroy}, C. \& {Spergel}, D.~N. 2011, \apj, 726, 36

\bibitem[{{Dalessandro} {et~al.}(2011){Dalessandro}, {Salaris}, {Ferraro},
  {Cassisi}, {Lanzoni}, {Rood}, {Fusi Pecci}, \& {Sabbi}}]{dalessandro:11}
{Dalessandro}, E., {Salaris}, M., {Ferraro}, F.~R., {et~al.} 2011, \mnras, 410,
  694

\bibitem[{{Dalessandro} {et~al.}(2013){Dalessandro}, {Salaris}, {Ferraro},
  {Mucciarelli}, \& {Cassisi}}]{dalessandro:13}
{Dalessandro}, E., {Salaris}, M., {Ferraro}, F.~R., {Mucciarelli}, A., \&
  {Cassisi}, S. 2013, \mnras, 430, 459

\bibitem[{{D'Antona} {et~al.}(2002){D'Antona}, {Caloi}, {Montalb{\'a}n},
  {Ventura}, \& {Gratton}}]{dantona:02}
{D'Antona}, F., {Caloi}, V., {Montalb{\'a}n}, J., {Ventura}, P., \& {Gratton},
  R. 2002, \aap, 395, 69

\bibitem[{{de Mink} {et~al.}(2009){de Mink}, {Pols}, {Langer}, \&
  {Izzard}}]{demink}
{de Mink}, S.~E., {Pols}, O.~R., {Langer}, N., \& {Izzard}, R.~G. 2009, \aap,
  507, L1

\bibitem[{{Decressin} {et~al.}(2007){Decressin}, {Meynet}, {Charbonnel},
  {Prantzos}, \& {Ekstr{\"o}m}}]{decressin}
{Decressin}, T., {Meynet}, G., {Charbonnel}, C., {Prantzos}, N., \&
  {Ekstr{\"o}m}, S. 2007, \aap, 464, 1029

\bibitem[{{Denissenkov} {et~al.}(2014){Denissenkov}, {VandenBerg}, {Hartwick},
  {Herwig}, {Weiss}, \& {Paxton}}]{denissenkov:14}
{Denissenkov}, P., {VandenBerg}, D., {Hartwick}, D., {et~al.} 2014, ArXiv
  e-prints

\bibitem[{{D'Ercole} {et~al.}(2011){D'Ercole}, {D'Antona}, \&
  {Vesperini}}]{dercole:11}
{D'Ercole}, A., {D'Antona}, F., \& {Vesperini}, E. 2011, \mnras, 415, 1304

\bibitem[{{Dorman} {et~al.}(1993){Dorman}, {Rood}, \& {O'Connell}}]{dorman:93}
{Dorman}, B., {Rood}, R.~T., \& {O'Connell}, R.~W. 1993, \apj, 419, 596

\bibitem[{{Gabriel} {et~al.}(2014){Gabriel}, {Noels}, {Montalban}, \&
  {Miglio}}]{gabriel}
{Gabriel}, M., {Noels}, A., {Montalban}, J., \& {Miglio}, A. 2014, ArXiv
  e-prints

\bibitem[{{Gratton} {et~al.}(2003){Gratton}, {Bragaglia}, {Carretta},
  {Clementini}, {Desidera}, {Grundahl}, \& {Lucatello}}]{gbc03}
{Gratton}, R.~G., {Bragaglia}, A., {Carretta}, E., {et~al.} 2003, \aap, 408,
  529

\bibitem[{{Gratton} {et~al.}(2012){Gratton}, {Carretta}, \&
  {Bragaglia}}]{gratton:12}
{Gratton}, R.~G., {Carretta}, E., \& {Bragaglia}, A. 2012, \aapr, 20, 50

\bibitem[{{Gratton} {et~al.}(2010){Gratton}, {D'Orazi}, {Bragaglia},
  {Carretta}, \& {Lucatello}}]{gdbcl:10}
{Gratton}, R.~G., {D'Orazi}, V., {Bragaglia}, A., {Carretta}, E., \&
  {Lucatello}, S. 2010, \aap, 522, A77

\bibitem[{{Gratton} {et~al.}(2011){Gratton}, {Lucatello}, {Carretta},
  {Bragaglia}, {D'Orazi}, \& {Momany}}]{gratton:11}
{Gratton}, R.~G., {Lucatello}, S., {Carretta}, E., {et~al.} 2011, \aap, 534,
  A123

\bibitem[{{Greggio} \& {Renzini}(1990)}]{gr:90}
{Greggio}, L. \& {Renzini}, A. 1990, \apj, 364, 35

\bibitem[{{Grundahl} {et~al.}(1999){Grundahl}, {Catelan}, {Landsman},
  {Stetson}, \& {Andersen}}]{grundahl:99}
{Grundahl}, F., {Catelan}, M., {Landsman}, W.~B., {Stetson}, P.~B., \&
  {Andersen}, M.~I. 1999, \apj, 524, 242

\bibitem[{{Hu} {et~al.}(2011){Hu}, {Tout}, {Glebbeek}, \& {Dupret}}]{hu:11}
{Hu}, H., {Tout}, C.~A., {Glebbeek}, E., \& {Dupret}, M.-A. 2011, \mnras, 418,
  195

\bibitem[{{Johnson} \& {Pilachowski}(2012)}]{jp:12}
{Johnson}, C.~I. \& {Pilachowski}, C.~A. 2012, \apjl, 754, L38

\bibitem[{{Marino} {et~al.}(2014){Marino}, {Milone}, {Przybilla}, {Bergemann},
  {Lind}, {Asplund}, {Cassisi}, {Catelan}, {Casagrande}, {Valcarce}, {Bedin},
  {Cort{\'e}s}, {D'Antona}, {Jerjen}, {Piotto}, {Schlesinger}, {Zoccali}, \&
  {Angeloni}}]{marino:14}
{Marino}, A.~F., {Milone}, A.~P., {Przybilla}, N., {et~al.} 2014, \mnras, 437,
  1609

\bibitem[{{Marino} {et~al.}(2012){Marino}, {Milone}, {Sneden}, {Bergemann},
  {Kraft}, {Wallerstein}, {Cassisi}, {Aparicio}, {Asplund}, {Bedin}, {Hilker},
  {Lind}, {Momany}, {Piotto}, {Roederer}, {Stetson}, \& {Zoccali}}]{marino:12}
{Marino}, A.~F., {Milone}, A.~P., {Sneden}, C., {et~al.} 2012, \aap, 541, A15

\bibitem[{{Marino} {et~al.}(2011){Marino}, {Villanova}, {Milone}, {Piotto},
  {Lind}, {Geisler}, \& {Stetson}}]{marino:11}
{Marino}, A.~F., {Villanova}, S., {Milone}, A.~P., {et~al.} 2011, \apjl, 730,
  L16

\bibitem[{{Michaud} {et~al.}(2008){Michaud}, {Richer}, \&
  {Richard}}]{michaud:08}
{Michaud}, G., {Richer}, J., \& {Richard}, O. 2008, \apj, 675, 1223

\bibitem[{{Michaud} {et~al.}(2011){Michaud}, {Richer}, \&
  {Richard}}]{michaud:11}
{Michaud}, G., {Richer}, J., \& {Richard}, O. 2011, \aap, 529, A60

\bibitem[{{Milone} {et~al.}(2014){Milone}, {Marino}, {Dotter}, {Norris},
  {Jerjen}, {Piotto}, {Cassisi}, {Bedin}, {Recio Blanco}, {Sarajedini},
  {Asplund}, {Monelli}, \& {Aparicio}}]{milone:14}
{Milone}, A.~P., {Marino}, A.~F., {Dotter}, A., {et~al.} 2014, \apj, 785, 21

\bibitem[{{Milone} {et~al.}(2013){Milone}, {Marino}, {Piotto}, {Bedin},
  {Anderson}, {Aparicio}, {Bellini}, {Cassisi}, {D'Antona}, {Grundahl},
  {Monelli}, \& {Yong}}]{mmp:13}
{Milone}, A.~P., {Marino}, A.~F., {Piotto}, G., {et~al.} 2013, \apj, 767, 120

\bibitem[{{Moehler} {et~al.}(2004){Moehler}, {Sweigart}, {Landsman}, {Hammer},
  \& {Dreizler}}]{Moh04}
{Moehler}, S., {Sweigart}, A.~V., {Landsman}, W.~B., {Hammer}, N.~J., \&
  {Dreizler}, S. 2004, \aap, 415, 313

\bibitem[{{Momany} {et~al.}(2002){Momany}, {Piotto}, {Recio-Blanco}, {Bedin},
  {Cassisi}, \& {Bono}}]{momany:02}
{Momany}, Y., {Piotto}, G., {Recio-Blanco}, A., {et~al.} 2002, \apjl, 576, L65

\bibitem[{{Norris}(1981)}]{norris}
{Norris}, J. 1981, \apj, 248, 177

\bibitem[{{Norris} {et~al.}(1981){Norris}, {Cottrell}, {Freeman}, \& {Da
  Costa}}]{norris:81}
{Norris}, J., {Cottrell}, P.~L., {Freeman}, K.~C., \& {Da Costa}, G.~S. 1981,
  \apj, 244, 205

\bibitem[{{Pace} {et~al.}(2006){Pace}, {Recio-Blanco}, {Piotto}, \&
  {Momany}}]{pace:06}
{Pace}, G., {Recio-Blanco}, A., {Piotto}, G., \& {Momany}, Y. 2006, \aap, 452,
  493

\bibitem[{{Pietrinferni} {et~al.}(2004){Pietrinferni}, {Cassisi}, {Salaris}, \&
  {Castelli}}]{basti1}
{Pietrinferni}, A., {Cassisi}, S., {Salaris}, M., \& {Castelli}, F. 2004, \apj,
  612, 168

\bibitem[{{Pietrinferni} {et~al.}(2006){Pietrinferni}, {Cassisi}, {Salaris}, \&
  {Castelli}}]{basti2}
{Pietrinferni}, A., {Cassisi}, S., {Salaris}, M., \& {Castelli}, F. 2006, \apj,
  642, 797

\bibitem[{{Pietrinferni} {et~al.}(2009){Pietrinferni}, {Cassisi}, {Salaris},
  {Percival}, \& {Ferguson}}]{basti:09}
{Pietrinferni}, A., {Cassisi}, S., {Salaris}, M., {Percival}, S., \&
  {Ferguson}, J.~W. 2009, \apj, 697, 275

\bibitem[{{Piotto} {et~al.}(2012){Piotto}, {Milone}, {Anderson}, {Bedin},
  {Bellini}, {Cassisi}, {Marino}, {Aparicio}, \& {Nascimbeni}}]{piotto:12}
{Piotto}, G., {Milone}, A.~P., {Anderson}, J., {et~al.} 2012, \apj, 760, 39

\bibitem[{{Reimers}(1975)}]{reimers}
{Reimers}, D. 1975, Memoires of the Societe Royale des Sciences de Liege, 8,
  369

\bibitem[{{Renzini} {et~al.}(1996){Renzini}, {Bragaglia}, {Ferraro},
  {Gilmozzi}, {Ortolani}, {Holberg}, {Liebert}, {Wesemael}, \&
  {Bohlin}}]{rbf96}
{Renzini}, A., {Bragaglia}, A., {Ferraro}, F.~R., {et~al.} 1996, \apjl, 465,
  L23

\bibitem[{{Rood}(1973)}]{r1973}
{Rood}, R.~T. 1973, \apj, 184, 815

\bibitem[{{Salaris} {et~al.}(2002){Salaris}, {Cassisi}, \& {Weiss}}]{scw}
{Salaris}, M., {Cassisi}, S., \& {Weiss}, A. 2002, \pasp, 114, 375

\bibitem[{{Sandquist}(2000)}]{sq00}
{Sandquist}, E.~L. 2000, \mnras, 313, 571

\bibitem[{{Sandquist} \& {Bolte}(2004)}]{sb:04}
{Sandquist}, E.~L. \& {Bolte}, M. 2004, \apj, 611, 323

\bibitem[{{Sbordone} {et~al.}(2011){Sbordone}, {Salaris}, {Weiss}, \&
  {Cassisi}}]{sbordone:11}
{Sbordone}, L., {Salaris}, M., {Weiss}, A., \& {Cassisi}, S. 2011, \aap, 534,
  A9

\bibitem[{{Smith} \& {Norris}(1993)}]{sn:93}
{Smith}, G.~H. \& {Norris}, J.~E. 1993, \aj, 105, 173

\bibitem[{{Valcarce} \& {Catelan}(2011)}]{valcarce:11}
{Valcarce}, A.~A.~R. \& {Catelan}, M. 2011, \aap, 533, A120

\bibitem[{{Villanova} {et~al.}(2009){Villanova}, {Piotto}, \&
  {Gratton}}]{villanova:09}
{Villanova}, S., {Piotto}, G., \& {Gratton}, R.~G. 2009, \aap, 499, 755

\bibitem[{{Vink} \& {Cassisi}(2002)}]{vc:02}
{Vink}, J.~S. \& {Cassisi}, S. 2002, \aap, 392, 553

\bibitem[{{Weiss} \& {Ferguson}(2009)}]{wf:09}
{Weiss}, A. \& {Ferguson}, J.~W. 2009, \aap, 508, 1343

\bibitem[{{Yong} {et~al.}(2013){Yong}, {Mel{\'e}ndez}, {Grundahl}, {Roederer},
  {Norris}, {Milone}, {Marino}, {Coelho}, {McArthur}, {Lind}, {Collet}, \&
  {Asplund}}]{ymg13}
{Yong}, D., {Mel{\'e}ndez}, J., {Grundahl}, F., {et~al.} 2013, \mnras, 434,
  3542

\bibitem[{{Yong} {et~al.}(2000){Yong}, {Demarque}, \& {Yi}}]{yong:00}
{Yong}, H., {Demarque}, P., \& {Yi}, S. 2000, \apj, 539, 928

\end{thebibliography}

\end{document}